\documentclass[12pt]{article}

\usepackage{graphicx}

\input{epsf}

\advance \topmargin by -\headheight
\advance \topmargin by -\headsep

\evensidemargin \oddsidemargin
\begin{document}

\topmargin -.6in

\def\rh{{\hat \rho}}
\def\alie{{\hat{\cal G}}}
\newcommand{\sect}[1]{\setcounter{equation}{0}\section{#1}}
\renewcommand{\theequation}{\thesection.\arabic{equation}}

\def\rf#1{(\ref{eq:#1})}
\def\lab#1{\label{eq:#1}}
\def\nonu{\nonumber}
\def\br{\begin{eqnarray}}
\def\er{\end{eqnarray}}
\def\be{\begin{equation}}
\def\ee{\end{equation}}
\def\eq{\!\!\!\! &=& \!\!\!\! }
\def\foot#1{\footnotemark\footnotetext{#1}}
\def\lb{\lbrack}
\def\rb{\rbrack}
\def\llangle{\left\langle}
\def\rrangle{\right\rangle}
\def\blangle{\Bigl\langle}
\def\brangle{\Bigr\rangle}
\def\llbrack{\left\lbrack}
\def\rrbrack{\right\rbrack}
\def\lcurl{\left\{}
\def\rcurl{\right\}}
\def\({\left(}
\def\){\right)}
\newcommand{\nit}{\noindent}
\newcommand{\ct}[1]{\cite{#1}}
\newcommand{\bi}[1]{\bibitem{#1}}
\def\lskip{\vskip\baselineskip\vskip-\parskip\noindent}
\relax

\def\tr{\mathop{\rm tr}}
\def\Tr{\mathop{\rm Tr}}
\def\v{\vert}
\def\bv{\bigm\vert}
\def\Bgv{\;\Bigg\vert}
\def\bgv{\bigg\vert}
\newcommand\partder[2]{{{\partial {#1}}\over{\partial {#2}}}}
\newcommand\funcder[2]{{{\delta {#1}}\over{\delta {#2}}}}
\newcommand\Bil[2]{\Bigl\langle {#1} \Bigg\vert {#2} \Bigr\rangle}  
\newcommand\bil[2]{\left\langle {#1} \bigg\vert {#2} \right\rangle} 
\newcommand\me[2]{\left\langle {#1}\bv {#2} \right\rangle} 
\newcommand\sbr[2]{\left\lbrack\,{#1}\, ,\,{#2}\,\right\rbrack}
\newcommand\pbr[2]{\{\,{#1}\, ,\,{#2}\,\}}
\newcommand\pbbr[2]{\lcurl\,{#1}\, ,\,{#2}\,\rcurl}

\def\ket#1{\mid {#1} \rangle}
\def\bra#1{\langle {#1} \mid}
\newcommand{\braket}[2]{\langle {#1} \mid {#2}\rangle}
%
\def\a{\alpha}
\def\at{{\tilde A}^R}
\def\atc{{\tilde {\cal A}}^R}
\def\atcm#1{{\tilde {\cal A}}^{(R,#1)}}
\def\b{\beta}
\def\dc{{\cal D}}
\def\d{\delta}
\def\D{\Delta}
\def\eps{\epsilon}
\def\vareps{\varepsilon}
\def\g{\gamma}
\def\G{\Gamma}
\def\grad{\nabla}
\def\h{{1\over 2}}
\def\l{\lambda}
\def\L{\Lambda}
\def\m{\mu}
\def\n{\nu}
\def\o{\over}
\def\om{\omega}
\def\O{\Omega}
\def\p{\phi}
\def\P{\Phi}
\def\pa{\partial}
\def\pr{\prime}
\def\pt{{\tilde \Phi}}
\def\qs{Q_{\bf s}}
\def\ra{\rightarrow}
\def\s{\sigma}
\def\S{\Sigma}
\def\t{\tau}
\def\th{\theta}
\def\Th{\Theta}
\def\tpp{\Theta_{+}}
\def\tmm{\Theta_{-}}
\def\tpg{\Theta_{+}^{>}}
\def\tms{\Theta_{-}^{<}}
\def\tp0{\Theta_{+}^{(0)}}
\def\tm0{\Theta_{-}^{(0)}}
\def\ti{\tilde}
\def\wti{\widetilde}
\def\jc{J^C}
\def\bj{{\bar J}}
\def\sj{{\jmath}}
\def\bsj{{\bar \jmath}}
\def\bp{{\bar \p}}
\def\vp{\varphi}
\def\ve{\varepsilon}
\def\vt{{\tilde \varphi}}
\def\faa{Fa\'a di Bruno~}
\def\ca{{\cal A}}
\def\cb{{\cal B}}
\def\ce{{\cal E}}
\def\cg{{\cal G}}
\def\cgh{{\hat {\cal G}}}
\def\ch{{\cal H}}
\def\chh{{\hat {\cal H}}}
\def\cl{{\cal L}}
\def\cm{{\cal M}}
\def\cn{{\cal N}}
\newcommand\sumi[1]{\sum_{#1}^{\infty}}   
\newcommand\fourmat[4]{\left(\begin{array}{cc}  
{#1} & {#2} \\ {#3} & {#4} \end{array} \right)}

%
\def\lie{{\cal G}}
\def\kmlie{{\hat{\cal G}}}
\def\dlie{{\cal G}^{\ast}}
\def\elie{{\widetilde \lie}}
\def\edlie{{\elie}^{\ast}}
\def\hlie{{\cal H}}
\def\flie{{\cal F}}
\def\wlie{{\widetilde \lie}}
\def\f#1#2#3 {f^{#1#2}_{#3}}
\def\winf{{\sf w_\infty}}
\def\win1{{\sf w_{1+\infty}}}
\def\hwinf{{\sf {\hat w}_{\infty}}}
\def\Winf{{\sf W_\infty}}
\def\Win1{{\sf W_{1+\infty}}}
\def\hWinf{{\sf {\hat W}_{\infty}}}
\def\Rm#1#2{r(\vec{#1},\vec{#2})}          
\def\OR#1{{\cal O}(R_{#1})}           
\def\ORti{{\cal O}({\widetilde R})}           
\def\AdR#1{Ad_{R_{#1}}}              
\def\dAdR#1{Ad_{R_{#1}^{\ast}}}      
\def\adR#1{ad_{R_{#1}^{\ast}}}       
\def\KP{${\rm \, KP\,}$}                 
\def\KPl{${\rm \,KP}_{\ell}\,$}         
\def\KPo{${\rm \,KP}_{\ell = 0}\,$}         
\def\mKPa{${\rm \,KP}_{\ell = 1}\,$}    
\def\mKPb{${\rm \,KP}_{\ell = 2}\,$}    
%
\def\rlx{\relax\leavevmode}
\def\inbar{\vrule height1.5ex width.4pt depth0pt}
\def\IZ{\rlx\hbox{\sf Z\kern-.4em Z}}
\def\IR{\rlx\hbox{\rm I\kern-.18em R}}
\def\IC{\rlx\hbox{\,$\inbar\kern-.3em{\rm C}$}}
\def\IN{\rlx\hbox{\rm I\kern-.18em N}}
\def\IO{\rlx\hbox{\,$\inbar\kern-.3em{\rm O}$}}
\def\IP{\rlx\hbox{\rm I\kern-.18em P}}
\def\IQ{\rlx\hbox{\,$\inbar\kern-.3em{\rm Q}$}}
\def\IF{\rlx\hbox{\rm I\kern-.18em F}}
\def\IG{\rlx\hbox{\,$\inbar\kern-.3em{\rm G}$}}
\def\IH{\rlx\hbox{\rm I\kern-.18em H}}
\def\II{\rlx\hbox{\rm I\kern-.18em I}}
\def\IK{\rlx\hbox{\rm I\kern-.18em K}}
\def\IL{\rlx\hbox{\rm I\kern-.18em L}}
\def\one{\hbox{{1}\kern-.25em\hbox{l}}}
\def\0#1{\relax\ifmmode\mathaccent"7017{#1}%
B        \else\accent23#1\relax\fi}
\def\omz{\0 \omega}
%
\def\ltimes{\mathrel{\vrule height1ex}\joinrel\mathrel\times}
\def\rtimes{\mathrel\times\joinrel\mathrel{\vrule height1ex}}
%
\def\mark{\noindent{\bf Remark.}\quad}
\def\prop{\noindent{\bf Proposition.}\quad}
\def\theor{\noindent{\bf Theorem.}\quad}
\def\name{\noindent{\bf Definition.}\quad}
\def\exam{\noindent{\bf Example.}\quad}
\def\proof{\noindent{\bf Proof.}\quad}
%
%
\def\PRL#1#2#3{{\sl Phys. Rev. Lett.} {\bf#1} (#2) #3}
\def\NPB#1#2#3{{\sl Nucl. Phys.} {\bf B#1} (#2) #3}
\def\NPBFS#1#2#3#4{{\sl Nucl. Phys.} {\bf B#2} [FS#1] (#3) #4}
\def\CMP#1#2#3{{\sl Commun. Math. Phys.} {\bf #1} (#2) #3}
\def\PRD#1#2#3{{\sl Phys. Rev.} {\bf D#1} (#2) #3}
\def\PLA#1#2#3{{\sl Phys. Lett.} {\bf #1A} (#2) #3}
\def\PLB#1#2#3{{\sl Phys. Lett.} {\bf #1B} (#2) #3}
\def\JMP#1#2#3{{\sl J. Math. Phys.} {\bf #1} (#2) #3}
\def\PTP#1#2#3{{\sl Prog. Theor. Phys.} {\bf #1} (#2) #3}
\def\SPTP#1#2#3{{\sl Suppl. Prog. Theor. Phys.} {\bf #1} (#2) #3}
\def\AoP#1#2#3{{\sl Ann. of Phys.} {\bf #1} (#2) #3}
\def\PNAS#1#2#3{{\sl Proc. Natl. Acad. Sci. USA} {\bf #1} (#2) #3}
\def\RMP#1#2#3{{\sl Rev. Mod. Phys.} {\bf #1} (#2) #3}
\def\PR#1#2#3{{\sl Phys. Reports} {\bf #1} (#2) #3}
\def\AoM#1#2#3{{\sl Ann. of Math.} {\bf #1} (#2) #3}
\def\UMN#1#2#3{{\sl Usp. Mat. Nauk} {\bf #1} (#2) #3}
\def\FAP#1#2#3{{\sl Funkt. Anal. Prilozheniya} {\bf #1} (#2) #3}
\def\FAaIA#1#2#3{{\sl Functional Analysis and Its Application} {\bf #1} (#2)
#3}
\def\BAMS#1#2#3{{\sl Bull. Am. Math. Soc.} {\bf #1} (#2) #3}
\def\TAMS#1#2#3{{\sl Trans. Am. Math. Soc.} {\bf #1} (#2) #3}
\def\InvM#1#2#3{{\sl Invent. Math.} {\bf #1} (#2) #3}
\def\LMP#1#2#3{{\sl Letters in Math. Phys.} {\bf #1} (#2) #3}
\def\IJMPA#1#2#3{{\sl Int. J. Mod. Phys.} {\bf A#1} (#2) #3}
\def\AdM#1#2#3{{\sl Advances in Math.} {\bf #1} (#2) #3}
\def\RMaP#1#2#3{{\sl Reports on Math. Phys.} {\bf #1} (#2) #3}
\def\IJM#1#2#3{{\sl Ill. J. Math.} {\bf #1} (#2) #3}
\def\APP#1#2#3{{\sl Acta Phys. Polon.} {\bf #1} (#2) #3}
\def\TMP#1#2#3{{\sl Theor. Mat. Phys.} {\bf #1} (#2) #3}
\def\JPA#1#2#3{{\sl J. Physics} {\bf A#1} (#2) #3}
\def\JSM#1#2#3{{\sl J. Soviet Math.} {\bf #1} (#2) #3}
\def\MPLA#1#2#3{{\sl Mod. Phys. Lett.} {\bf A#1} (#2) #3}
\def\JETP#1#2#3{{\sl Sov. Phys. JETP} {\bf #1} (#2) #3}
\def\JETPL#1#2#3{{\sl  Sov. Phys. JETP Lett.} {\bf #1} (#2) #3}
\def\PHSA#1#2#3{{\sl Physica} {\bf A#1} (#2) #3}
\def\PHSD#1#2#3{{\sl Physica} {\bf D#1} (#2) #3}
\def\PJA#1#2#3{{\sl Proc. Japan. Acad} {\bf #1A} (#2) #3}
\def\JPSJ#1#2#3{{\sl J. Phys. Soc. Japan} {\bf #1} (#2) #3}

\begin{titlepage}
\vspace*{-1cm}

\vskip 3cm

\vspace{.2in}
\begin{center}
{\large\bf A Skyrme-like model with an exact BPS bound}
\end{center}

\vspace{.5cm}

\begin{center}
L. A. Ferreira~$^{\star}$ and Wojtek J. Zakrzewski~$^{\dagger}$

\vspace{.3 in}
\small

\par \vskip .2in \noindent
$^{(\star)}$Instituto de F\'\i sica de S\~ao Carlos; IFSC/USP;\\
Universidade de S\~ao Paulo  \\ 
Caixa Postal 369, CEP 13560-970, S\~ao Carlos-SP, Brazil\\

\par \vskip .2in \noindent
$^{(\dagger)}$~Department of Mathematical Sciences,\\
 University of Durham, Durham DH1 3LE, U.K.

\normalsize
\end{center}

\vspace{.5in}

\begin{abstract}
 
We propose a new Skyrme-like model with fields taking values on the sphere $S^3$ or, equivalently, on the group $SU(2)$.  The action of the model contains  a quadratic kinetic term plus a quartic term  which is the same as that   of the Skyrme-Faddeev model. The novelty of the model is that it possess a first order Bogomolny type equation whose solutions automatically satisfy the second order Euler-Lagrange equations. It also possesses a lower bound on the static energy which is saturated by the Bogomolny solutions. Such Bogomolny equation is equivalent to the so-called force free equation used in plasma and solar Physics, and which possesses  large classes of solutions.  An old result due to Chandrasekhar prevents the existence of finite energy solutions for the force free equation on the entire tridimensional space $\IR^3$. We construct new exact finite energy solutions to the Bogomolny equations for the case where the space is  the three-sphere $S^3$, using  toroidal like coordinates.

\end{abstract} 
\end{titlepage}

\section{Introduction}
\label{sec:intro}
\setcounter{equation}{0}

One of the most important concepts used in the study of topological solitons is that of self-duality. The instantons in four dimensional Euclidean space and the so-called BPS monopoles and dyons, static solutions in four dimensional Minkowski space-time, are the best known examples of self-dual solutions. They are solutions of self-dual or Bogomolny  first order differential equations which automatically satisfy the second-order equations of motion. In addition,  they saturate lower bounds on the Euclidean action; in the case of instantons, and on the static energy in the case of BPS monopoles and dyons. The Skyrme \cite{skyrme} and Skyrme-Faddeev \cite{sf} models are important four dimensional field theories possessing topological solitons \cite{numericalsf} with a wide variety of applications in several areas of physics. However, such theories do not possess  first order Bogomolny type equations, nor an exact lower bound on their static energy. 
 They  possess instead inequalities that the energy functional has to satisfy. These inequalities give an estimate (a lower bound) 
 of the lowest energy of the solitons of a given topological sector. The lack of the reduction to the first order equation  has certainly prevented the development of exact methods, like those used for instantons and BPS monopoles. Recently some attempts have been made to modify the Skyrme theory to give it  a Bogomolny like equation, either by considering kinetic terms which are of order six in derivatives of the fields \cite{adam} or by coupling the original Skyrme model to an infinite tower of vector mesons \cite{paul}. 

In this paper we propose another Skyrme-like model that does possesses a self-dual sector. The fields of this model live on the sphere $S^3$, or equivalently on the Lie group $SU(2)$, and they can be parameterized by two complex scalar fields $Z_a$, $a=1,2$, satisfying the constraint $Z_a^*\,Z_a=1$. The action of the model is given by
\be
S= \int d^4x\( \frac{m^2}{2}\,A_{\mu}^2  - \frac{1}{4\, e^2}\, H_{\mu\nu}^2\)
\lab{ua845}
\ee
with $m$ and $e^2$ being coupling constants, of dimension of mass and dimensionless, respectively. In \rf{ua845} we have $\mu\, ,\,\nu=0,1,2,3$, and 
\be
A_{\mu}= \frac{i}{2}\(Z_a^*\partial_{\mu} Z_a-Z_a\partial_{\mu} Z_a^*\) \qquad\qquad {\rm and} \qquad\qquad
H_{\mu\nu}=\partial_{\mu} A_{\nu}-\partial_{\nu} A_{\mu}.
\lab{adef}
\ee

 In this paper we consider this model for two types of space-times: the four dimensional Minkowski space and the Einstein space $S^3\times \IR$. The Bogomolny or self-duality equation for static solutions of the model is given by
 \be
 B_i=\pm \, m\,e\,A_i \qquad\qquad {\rm with} \qquad \qquad B_i=\frac{1}{2} \, \varepsilon_{ijk}\, H_{jk} \qquad i,j,k=1,2,3
 \lab{selfdualeq}
\ee
and the solutions of these equations automatically satisfy the second-order Euler-Lagrange equations associated to \rf{ua845}. The static energy has a lower bound 
\be
E \geq 4\,\pi^2\,\frac{m}{e}\,\mid Q\mid
\lab{bound}
\ee
with $Q$ being the winding number of the map $S^3_S\rightarrow S^3_T$, where $S^3_T$ is the target space and $S^3_S$ is the spatial submanifold of the space-time. In the case of the Minkowski space-time the requirement of finite energy imposes the condition that the fields are constant at spatial infinity and so $\IR^3$ can be compactified into $S^3$, as far as homotopy issues are concerned. The solutions of \rf{selfdualeq} saturate the bound \rf{bound}. 

The paper is organized as follows. In section \ref{sec:model} we discuss the self-duality ideas leading to the construction of the proposed Skyrme model and present its main properties.  In section \ref{sec:virial} we discuss the connections of our Bogomolny equations with the so-called force-free equations used in plasma and solar physics, and describe the difficulties in constructing finite energy solutions in the entire space $\IR^3$. Finally, in section \ref{sec:sphere} we look at our model on the three sphere $S^3$ and construct its finite energy solutions for its  Bogomolny equations. 

\section{The construction of the model}
\label{sec:model}
\setcounter{equation}{0}

The self-duality or Bogomolny type equations have two striking features. Firstly, they are first order differential equations which imply the second order Euler-Lagrange equations, and secondly they lead to a  lower bound on the static energy and their solutions saturate that bound. The one less integration one has to perform to construct their solutions does not come, as usual, by the use of dynamically conserved quantities. Instead it comes
 from the invariance of a topological charge under smooth deformations of the fields, as explained in \cite{selfdual}. In the case of the model under consideration the relevant topological charge is the winding number of the map $S^3_S\rightarrow S^3_T$, provided by the solutions from the the spatial submanifold $S^3_S$ to the target space $S^3_T$. Such a winding number is given by the integral formula 
\be
Q=\frac{1}{12\,\pi^2}\,\int d^3x\, \varepsilon_{abcd}\,\varepsilon_{ijk}\,\Phi_{a}\,\partial_i\Phi_{b}\,\partial_j\Phi_{c}\,\partial_k \Phi_{d}=\frac{1}{4\,\pi^2}\,\int d^3 x\, A_i\,B_i,
\lab{topcharge}
\ee
where we have written $Z_1\equiv \Phi_1+i\,\Phi_2$, $Z_2\equiv \Phi_3+i\,\Phi_4$, and $a,b,c,d=1,2,3,4$. Note that on the r.h.s. of \rf{topcharge} we have written $Q$ in terms of the vectors $A_i$ and $B_i$ defined in \rf{adef} and \rf{selfdualeq}, respectively, and it reminds the Hopf invariant used in the theories with the target space being $S^2$, like in the the Skyrme-Faddeev model. However, our target space is $S^3_T$ and we are not projecting the map down to $S^2$ as is the case for the Hopf map. 

Next we follow the arguments presented in \cite{selfdual}. Let us denote by $\zeta_{\alpha}$, $\alpha=1,2,3$, the independent fields of the target space $S^3_T$. The topological charge $Q$ given in  \rf{topcharge} is invariant under infinitesimal smooth (homotopic) deformations of the fields $\delta \zeta_{\alpha}$, and so, without the use of the equations of motion, one finds that $\delta Q=0$. Since the variations are arbitrary one gets from \rf{topcharge} that the vectors $A_i$ and $B_i$ have to satisfy
\be
B_i\frac{\delta A_i}{\delta \zeta_{\alpha}}
-  \partial_{j}\(B_i\frac{\delta A_i}{\delta \partial_{j}\zeta_{\alpha}}\)
+A_i\frac{\delta B_i}{\delta \zeta_{\alpha}}
-  \partial_{j}\(A_i\frac{\delta B_i}{\delta \partial_{j}\zeta_{\alpha}}\)=0 .
\lab{identity}
\ee

On the other hand, the static Euler-Lagrange equations coming from \rf{ua845} are given by 
\be
m^2\,e^2\, A_i\frac{\delta A_i}{\delta \zeta_{\alpha}}
-  m^2\,e^2\, \partial_{j}\(A_i\frac{\delta A_i}{\delta \partial_{j}\zeta_{\alpha}}\)
+B_i\frac{\delta B_i}{\delta \zeta_{\alpha}}
-  \partial_{j}\(B_i\frac{\delta B_i}{\delta \partial_{j}\zeta_{\alpha}}\)=0 .
\lab{euler}
\ee

Thus one finds  that \rf{identity} and \rf{selfdualeq} together imply \rf{euler}. On the other hand, since \rf{identity} is an identity satisfied by any smooth field configuration, it follows that one has to solve just the first order equations \rf{selfdualeq} to find solutions of the second order Euler-Lagrange equations \rf{euler}.

Let us consider next the static energy of the model \rf{ua845}
\br
E&=& \frac{1}{2\,e^2}\,\int d^3x\( m^2\,e^2\,A_{i}^2  +  B_{i}^2\)
\nonumber\\
&=& \frac{1}{2\,e^2}\,\int d^3x\(B_i \pm  m\,e\,A_{i} \)^2+ 4\,\pi^2\,\frac{m}{e}\, \mid Q\mid
\er
Then one finds that the lower bound \rf{bound} indeed is satisfied and that the Bogomolny solutions of \rf{selfdualeq} saturate it. Thus the energy for such Bogomolny solutions can be written as
\be
E_{\rm Bogom.}= m^2\,\int d^3x\, A_i^2= \frac{1}{e^2}\,\int d^3x\, B_i^2=4\,\pi^2\,\frac{m}{e}\, \mid Q\mid.
\lab{selfdualenergy}
\ee

Note that the two terms in the energy functional balance each other according to Derrick's argument \cite{derrick}. Indeed, by rescaling the space coordinates as $x^i\rightarrow \lambda\, x^i$, one gets that 
 \br
E\(\lambda\)&=& \frac{1}{2}\,\int d^3x\( \lambda\, m^2\,A_{i}^2  +  \frac{1}{\lambda}\frac{1}{e^2}B_{i}^2\)\nonumber\\
&=& E + \(\lambda -1\)\, \frac{1}{2}\,\int d^3x\(  m^2\,A_{i}^2  -  \frac{1}{e^2}B_{i}^2\)
+O\(\(\lambda -1\)^2\)
\er
Thus, the solutions of \rf{selfdualeq} automatically satisfy the condition of stability of Derrick's theorem. The problem, however, is that there may not exist any finite energy solutions of \rf{selfdualeq} (defined over the entire space $\IR^3$) as we will discuss in section \ref{sec:virial}.

\subsection{A parameterization of the fields}
\label{sec:fields}

We have the constraint $Z_a^*\, Z_a=1$ which we can solve using one complex scalar field $u$ and one real scalar field $\varphi$ by putting:
\be
Z_1= \frac{u\,e^{i\,\vp}}{\sqrt{1+\mid u\mid^2}}\qquad\qquad \qquad 
Z_2= \frac{e^{i\,\vp}}{\sqrt{1+\mid u\mid^2}}.
\lab{uvpdef}
\ee
In this parametrisation we find that $A_{\mu}$ and $H_{\mu\nu}$, introduced in \rf{adef}, becomes 
\be
A_{\mu}=  \frac{i}{2}\,\frac{\(u^*\,\partial_{\mu}u-u\,\partial_{\mu}u^*\)}{1+\mid u\mid^2} - \partial_{\mu}\vp
\qquad\qquad\qquad
H_{\mu\nu}=i\,\frac{\(\partial_{\mu}u^*\,\partial_{\nu}u-\partial_{\nu}u^*\,\partial_{\mu}u\)}{\(1+\mid u\mid^2\)^2}.
\lab{ahinuvp}
\ee

The Euler-Lagrange equation corresponding to \rf{ua845}, for the field $\vp$, is then given by
\be
\partial^{\mu}A_{\mu}=0 
\lab{eqvp}
\ee
and the corresponding equation for $u^*$,  using \rf{eqvp}, becomes
\be
\partial_{\mu}\(H^{\mu\nu}\,\partial_{\nu} u\) + m^2\, e^2\,A^{\mu}\,\partial_{\mu} u =0.
\lab{equ0}
\ee

The static Bogomolny equations \rf{selfdualeq} lead to the relations $B_i=\pm \, m\,e\,A_i=i\,\varepsilon_{ijk}\,\frac{\partial_{j}u^*\,\partial_{k}u}{\(1+\mid u\mid^2\)^2}$. Thus they imply that  $\partial_iA_i=0$ and $A_i\,\partial_i u=0$.  In addition, one finds that $\partial_{i}\(H_{ij}\,\partial_{j} u\)=\varepsilon_{ijk}\partial_{i}\(B_k\,\partial_{j} u\)=\pm\,m\,e\,\varepsilon_{ijk}\partial_{i}\(A_k\,\partial_{j} u\)=0$. So, the Bogomolny equations \rf{selfdualeq} do indeed imply the Euler-Lagrange equations \rf{eqvp} and \rf{equ0}.

\section{Solutions on $\IR^3$}
\label{sec:virial}
\setcounter{equation}{0}

The equations \rf{selfdualeq}, in the case where the vector ${\vec A}$ describes a magnetic field, has been extensively used in plasma and solar physics, and there is extensive literature about their solutions. An important result due to Chandrasekhar \cite{chandra} states that there cannot exist smooth finite energy solutions  to the equations  \rf{selfdualeq} in the entire space $\IR^3$. For completeness, we  reproduce here the arguments of Chandrasekhar, given on page 158 of his book \cite{chandra}, since they are quite important for our analysis. 

Let ${\vec C}$ be a vector field defined on the tridimensional space $\IR^3$, and let ${\vec r}$ be the position vector of a given point in that space. Then by simple and straightfoward manipulations one can derive the following identity:   
\be
{\vec r}\cdot \( {\rm curl} {\vec C}\wedge{\vec C}\)= \frac{1}{2}{\vec C}^2-\({\vec r}\cdot {\vec C}\)\,\({\vec \nabla}\cdot {\vec C}\)-\frac{\partial\,}{\partial x^i}\left[\frac{1}{2}x^i\,{\vec C}^2-\({\vec r}\cdot {\vec C}\)\, C^i\right].
\ee
If one now requires that the vector field satisfies
\be
{\rm curl} {\vec C}\wedge{\vec C}=0 \qquad\qquad \qquad {\vec \nabla}\cdot {\vec C}=0
\lab{cond}
\ee
it follows that
\be
{\vec C}^2 = {\vec \nabla}\cdot {\vec D} \qquad\qquad {\rm with}\qquad\qquad 
{\vec D}={\vec C}^2\,{\vec r}-2\,\({\vec r}\cdot {\vec C}\)\, {\vec C}.
\lab{nice}
\ee

The conditions \rf{cond} are equivalent to
\be
{\rm curl} {\vec C}=\alpha\, {\vec C}\qquad\qquad \qquad {\vec \nabla}\alpha \cdot {\vec C}=0,
\lab{forcefree}
\ee
which are known, in plasma and solar physics literature, as the equations for force free magnetic fields.  Indeed, if ${\vec C}$ is a magnetic field then in a plasma ${\rm curl} {\vec C}$ is proportional to the current density and so ${\rm curl} {\vec C}\wedge{\vec C}$ is proportional to the magnetic force. Since the magnetic forces in a plasma are much stronger than all other forces involved in the system it was suggested in \cite{taylor} that \rf{forcefree} are the conditions to obtain a relaxed state (see also \cite{wiegelmann}). 

Integrating \rf{nice} over the whole space $V\equiv \IR^3$ one gets 
\be
\int_V d^3x\, {\vec C}^2= \int_{\partial V} d{\vec S} \cdot\left[{\vec C}^2\,{\vec r}-2\,\({\vec r}\cdot {\vec C}\)\, {\vec C}\right],
\lab{virial}
\ee
which in the case when ${\vec C}$ is a magnetic field, is known as the virial theorem in magnetohydrodynamics (MHD). In order for the integral on the l.h.s. of \rf{virial} to be finite it is necessary that $ \mid {\vec C}\mid \rightarrow 1/r^{\(3+\varepsilon\)/2}$, as $r\rightarrow \infty$, for $\varepsilon>0$. However, this implies that the surface integral on the r.h.s. of \rf{virial} vanishes. So, the volume integral is finite only for  the trivial configuration. Thus the  force-free equations \rf{forcefree} cannot be satisfied over the entire space $\IR^3$ by a smooth finite energy  magnetic field ${\vec C}$.

Another argument, due to Seehafer \cite{seehafer}, states that there can be no finite energy solutions of \rf{forcefree}, for $\alpha$ constant, outside a compact three dimensional region (like  the Sun or any ball) in $\IR^3$.  There exist, however, two finite energy solutions of \rf{forcefree}, with constant $\alpha$, in non-compact regions. The first one exists in the case of a semi-infinite rectangular column; it was found by Seehafer \cite{seehafer}. The second one was constructed by Aly \cite{aly} in the case of a three dimensional half-space. An  extensive discussion of solutions of the force-free equations \rf{forcefree} can be found in the book by Marsh \cite{marsh}; most of them are defined inside regions of finite volume, like torus and spheres. 

The Bogomolny equations \rf{selfdualeq} correspond to the force-free equations \rf{forcefree} when we identify ${\vec A}$ with ${\vec C}$, and put $\alpha = \pm m\, e$. Therefore, all known solutions of the force free-equation can perhaps be solutions of our model. The finite energy solutions, however, will have to be nonzero only over a finite volume or over some special non-compact regions of $\IR^3$. Note, however, that  not all known solutions of \rf{forcefree} will be solutions of our model \rf{ua845}, since even when we know ${\vec A}$ we still have to find solutions for $Z_a$, or equivalently for $u$ and $\vp$ introduced in \rf{uvpdef}, by integrating  ${\vec A}$ given by \rf{adef} or equivalently by \rf{ahinuvp}. This is not always possible and we will discuss  in  the section \ref{sec:sphere} when this can be done and when it cannot in  the case of field configurations defined over the sphere $S^3$.


\section{Solutions on $S^3$}
\label{sec:sphere}
\setcounter{equation}{0}

Next we consider the problem on $S^3$. We show that we can construct new  finite energy solutions for the Bogomolny equations \rf{selfdualeq} on the sphere $S^3$, using toroidal like coordinates, by the embedding of $S^3$ into $\IR^4$ as in \cite{decarli,bonfim} 
{\it i.e.} we define
\br
x_1&=& r_0\, \sqrt{z}\,\cos \theta_2, \qquad \qquad x_3= r_0\, \sqrt{1-z}\,\cos \theta_1,
\nonumber\\
x_2&=& r_0\, \sqrt{z}\,\sin \theta_2, \qquad \qquad \, x_4= r_0\, \sqrt{1-z}\,\sin \theta_1,
\lab{coords3}
\er
where $r_0$ is the radius of $S^3$, and $0\leq z\leq 1$, $0\leq \theta_a\leq 2\,\pi$, $a=1,2$. The metric on $S^3$, in these coordinates,
 takes the form
\be
d\,s^2= r_0^2\left[ \frac{dz^2}{4\,z\,(1-z)}+(1-z)\,d\theta_1^2+ z\,d\theta_2^2\right]
\equiv h_z^2 dz^2+h_{\theta_1}^2 d\theta_1^2+h_{\theta_2}^2 d\theta_2^2.
\lab{metrics3}
\ee
In terms of $3$-vectors the equations \rf{selfdualeq} can be rewritten as 
\be
{\vec \nabla}\wedge {\vec A}=\pm\,m\,e\,{\vec A}
\lab{curleq}
\ee
 with  ${\vec A}= A_z\,{\vec e}_z + A_{\theta_1}\,{\vec e}_{\theta_1}+ A_{\theta_2}\,{\vec e}_{\theta_2}$. Here $\partial_{\zeta_i} {\vec r} = h_{\zeta_i}\, {\vec e}_{\zeta_i}$, for $\zeta_i=\{z\,,\,\theta_1\,,\,\theta_2\}$, with the scaling factors $h_{\zeta_i}$ defined in \rf{metrics3}. In components the equations \rf{curleq} become
 \br
 \frac{h_z}{h_{\theta_1}\,h_{\theta_2}}\,\left[\partial_{\theta_1} V_{\theta_2}-\partial_{\theta_2} V_{\theta_1}\right] &=& \pm\,m\,e\, V_z,
 \nonumber\\
 \frac{h_{\theta_1}}{h_{z}\,h_{\theta_2}}\,\left[\partial_{\theta_2} V_{z}-\partial_{z} V_{\theta_2}\right] &=& \pm\,m\,e\, V_{\theta_1},
\lab{curlcomp} \\
 \frac{h_{\theta_2}}{h_{z}\,h_{\theta_1}}\,\left[\partial_{z} V_{\theta_1}-\partial_{\theta_1} V_{z}\right] &=& \pm\,m\,e\, V_{\theta_2},
\nonumber
\er
where we have defined $V_{\zeta_i}\equiv h_{\zeta_i}\, A_{\zeta_i}$. We shall look for solutions which are independent of $\theta_1$ and $\theta_2$, and so the equations in \rf {curlcomp} become
\be
\lambda\, V_{\theta_1}\(z\)=-(1-z)\,\partial_zV_{\theta_2}\(z\),\qquad\qquad 
\lambda\, V_{\theta_2}\(z\)= z\,\partial_zV_{\theta_1}\(z\),\qquad\qquad V_z=0
\lab{simpleeq}
\ee
with $\lambda$ being a dimensionless parameter defined by 
\be
\lambda=\pm\,\frac{1}{2}\,m\,e\,r_0.
\lab{lambdadef}
\ee

 Note that if $V_{\theta_1}\(z\)$ is a solution of \rf{simpleeq} so is $V_{\theta_2}\(z\)=V_{\theta_1}\(1-z\)$, and vice-versa. So it follows that $V_{\theta_1}\(z\)$ has to satisfy the hypergeometric equation:
\be
z\,(1-z)\,V_{\theta_1}^{\prime\prime} + (1-z)\, V_{\theta_1}^{\prime}+\lambda^2\,V_{\theta_1}=0
\lab{aa}
\ee
{\it i.e.} $V_{\theta_1}= {\rm const.}\, F\(\lambda , - \lambda ; 1; z\)$ with  primes in \rf{aa} denoting the $z$-derivatives.

The energy \rf{selfdualenergy} of this solution becomes
\be
E_{\rm Bogom.} = 4\,\pi^2\,\frac{m}{e}\, \mid\lambda\mid\int_0^1dz\,\left[
\frac{V_{\theta_1}^2}{1-z}+\frac{V_{\theta_2}^2}{z}\right]
\lab{selfdualenergy2}
\ee
since the volume element on $S^3$, in the coordinates \rf{coords3}, is given by  $dv=\frac{r_0^3}{2}\,dz\,d\theta_1\,d\theta_2$. In order for the energy to be finite we need $V_{\theta_1}\rightarrow (1-z)^r$, $r\geq \frac{1}{2}$, for $z\rightarrow 1$, and $V_{\theta_2}\rightarrow z^s$, $s\geq \frac{1}{2}$, for $z\rightarrow 0$. 

However, we know that  $F(\alpha , \beta ; \gamma ; 1)=\frac{\Gamma(\gamma)\,\Gamma(\gamma-\alpha - \beta)}{\Gamma(\gamma-\alpha)\,\Gamma(\gamma - \beta)}$ (see 9.122.1 of \cite{grad}). Therefore, since $V_{\theta_1}= {\rm const.}\, F\(\lambda , - \lambda ; 1; z\)$ has to vanish at $z=1$, we need to require that 
\be
\lambda =  \pm 1\, , \pm 2\, , \pm 3\, , \ldots
\lab{goodvalues}
\ee
and $V_{\theta_1}$ and $V_{\theta_2}$ have to  be polynomials of degree  $\mid \lambda\mid $ in $z$, {\it i.e.}
\be
V_{\theta_1}=\beta\, \(1-z\)\, P_{\mid\lambda\mid}\(z\), \qquad\qquad\qquad
V_{\theta_2}=\beta\, z\, Q_{\mid\lambda\mid}\(z\)
\ee
with $\beta$ an arbitrary constant.  Here $P_{\mid\lambda\mid}$ and $Q_{\mid\lambda\mid}$ are given by
\br
P_{\mid\lambda\mid}\(z\) &=&\sum_{n=0}^{\mid\lambda\mid-1} a_n\,z^n\; ; \qquad\qquad\; \;\;
a_0=1\; ; 
\qquad\qquad a_n=(-1)^n \prod_{k=1}^n \frac{\lambda^2-k^2}{k^2}
\\
Q_{\mid\lambda\mid}\(z\) &=&-\mid\lambda\mid\, \sum_{n=0}^{\mid\lambda\mid-1} b_n\,z^n\; ; \quad\quad b_0=1\; ; 
\qquad\quad\;\; \;b_n=(-1)^n \prod_{k=1}^n \frac{\lambda^2-k^2}{k\,(k+1)}.
\nonumber
\er

It is not difficult to check that $P_{\mid\lambda\mid}$ and $Q_{\mid\lambda\mid}$ satisfy the orthogonality relations:
\be
\int_0^1dz\,(1-z)\, P_n\,P_m = \frac{1}{2\,n}\,\delta_{n,m}\; ; \qquad\qquad \qquad 
\int_0^1dz\, z\, Q_n\,Q_m =\frac{1}{2\,n}\,\delta_{n,m}.
\ee
In consequence, the energy \rf{selfdualenergy2} of such solutions is independent of $\lambda$, {\it i.e.} it is given by 
\be
E_{\rm Bogom.} = 4\,\pi^2\,\frac{m}{e}\, \beta^2.
\lab{selfdualenergy3}
\ee

The solution for the vector ${\vec A}$ then takes the form
\be
{\vec A}= \frac{\beta}{ r_0}\,
\left[ \sqrt{1-z}\, \,P_{\mid\lambda\mid}\(z\)\,{\vec e}_{\theta_1}+ \sqrt{z}\,\, Q_{\mid\lambda\mid}\(z\)\,{\vec e}_{\theta_2}\right].
\ee

Note, however, that such finite energy solutions are not, necessarily, finite energy solutions of the model \rf{ua845}. To see this we observe that, according to \rf{lambdadef} and \rf{goodvalues}, the coupling constants have to be tuned to the radius of the sphere $S^3$ as 
\be
m^2\,e^2= 4\, \frac{\lambda^2}{r_0^2} \qquad\qquad\qquad  \lambda =  \pm 1\, , \pm 2\, , \pm 3\, , \ldots
\ee
Moreover, so far we have solved the Bogomolny equation \rf{curleq} for the vector ${\vec A}$. We now have to find the corresponding solutions for the fields $Z_a$, $a=1,2$, or equivalently, for the fields $u$ and $\vp$. We will then see that only one value of $\lambda$ is permitted if we want to have a finite energy solution of \rf{ua845}.  

Let us parameterize the field $u$, introduced in \rf{uvpdef}, as
\be
u=\sqrt{\frac{g}{1-g}}\, e^{i\,\phi},\qquad\qquad \qquad 0\leq g\leq 1,\qquad\qquad 0\leq \phi\leq 2\,\pi.
\ee
Then the vector ${\vec A}$ given by \rf{ahinuvp}, in terms of these variables,  takes the form
\be
{\vec A}=- g\,{\vec \nabla}\phi-{\vec \nabla}\vp.
\lab{relauvp}
\ee

In terms of the solutions we have constructed for ${\vec A}$ the equations \rf{relauvp} take the form
\br
g\, \partial_z \phi + \partial_z \vp&=&0,
\nonumber\\
g\, \partial_{\theta_1} \phi + \partial_{\theta_1} \vp&=&- V_{\theta_1}\(z\),
\lab{gphivpeqs}\\
g\, \partial_{\theta_2} \phi + \partial_{\theta_2} \vp&=&- V_{\theta_2}\(z\).
\nonumber
\er

We shall try to solve these equations using a simple ansatz
\be
g\equiv g\(z\),\qquad\qquad\qquad  \phi= m_1\,\theta_1+m_2\,\theta_2,\qquad\qquad\qquad \vp= n_1\,\theta_1+n_2\,\theta_2
\ee
with $m_a$ and $n_a$, $a=1,2$, being integers in order to have the fields $Z_a$, $a=1,2$,  single valued on $S^3$. 
With this ansatz the first equation in \rf{gphivpeqs} is automatically satisfied and the other two require that
\be
g=-\frac{V_{\theta_1}+n_1}{m_1}=-\frac{V_{\theta_2}+n_2}{m_2}.
\ee

Using \rf{simpleeq} to eliminate $V_{\theta_2}$ we get the final equation for $V_{\theta_1}$:
\be
\frac{z}{\lambda}\partial_z V_{\theta_1} -\frac{m_2}{m_1}\,V_{\theta_1}-n_1\, \frac{m_2}{m_1}+n_2=0
\lab{badeq}
\ee

Note, however, that $V_{\theta_1}$ is a polynomial of degree $\mid \lambda\mid$ in $z$ and so the equation \rf{badeq} will be satisfied 
if we solve the  $\mid \lambda\mid+1$  algebraic relations for different powers of $z$ in \rf{badeq}. However, looking at the arbitrary power $z^k$, $k\neq 0$, we note that these relations require that  $\frac{k}{\lambda}-\frac{m_2}{m_1}=0$. Thus  the degree of $V_{\theta_1}$ must be unity, {\it i.e.} $\mid \lambda\mid=1$. In consequence we see that
\be
m_1=m_2, \qquad\qquad\qquad \qquad \beta=n_2-n_1
\ee
with the second relation coming from the term independent of $z$ in \rf{badeq}. Thus, the only permitted solution is of the form
\be
V_{\theta_1}=\(n_2-n_1\)\, \(1-z\)\,,  \qquad\qquad\qquad\qquad 
V_{\theta_2}=-\(n_2-n_1\)\, z\, 
\ee
and so
\be
g=-\frac{\left[n_1\,z+n_2\,(1-z)\right]}{m_1}.
\ee

Finally, we have to consider the boundary conditions for $g$. 
If we require that $g(0)=g(1)$ we have a trivial solution. So to get a non-trivial solution we have two acceptable boundary conditions:
\br
g\(0\)&=&0\; ;\qquad g\(1\)=1\; ;\qquad  \rightarrow \qquad n_2=0\; ;  \quad n_1=-m_1\qquad \rightarrow \qquad g\(z\)=z,
\nonumber\\
g\(0\)&=&1\; ;\qquad g\(1\)=0\; ;\qquad  \rightarrow \qquad n_1=0\; ;  \quad n_2=-m_1\qquad \rightarrow \qquad g\(z\)=1-z.
\nonumber
\er

Denoting $l\equiv m_1=m_2$,  we note that we have two nontrivial solutions, corresponding to the two choices of boundary conditions, 
\br
u&=& \sqrt{\frac{z}{1-z}}\, e^{i\,l\(\theta_1+\theta_2\)}, \qquad\qquad \qquad
Z_1= \sqrt{z}\, e^{i\,l\,\theta_2},
\lab{solution1}\\
\vp&=& - l\, \theta_1, \qquad\qquad\qquad\qquad \qquad\;\;\,
Z_2= \sqrt{1-z}\, e^{-i\,l\,\theta_1}
\nonumber
\er
and
\br
u&=& \sqrt{\frac{1-z}{z}}\, e^{i\,l\(\theta_1+\theta_2\)},\qquad\qquad \qquad
Z_1= \sqrt{1-z}\, e^{i\,l\,\theta_1}, 
\lab{solution2}\\
\vp&=& - l\, \theta_2, \qquad\qquad\qquad\qquad \qquad\;\;\,
Z_2= \sqrt{z}\, e^{-i\,l\,\theta_2}.
\nonumber
\er
In both cases the coupling constants have to be related to the radius of $S^3$ by
\be
m^2\,e^2=  \frac{4}{r_0^2}.
\ee

The energies of these solutions are equal and are given by 
\be
E_{\rm Bogom.} = 4\,\pi^2\,\frac{m}{e}\, l^2
\lab{selfdualenergy4}
\ee
and the corresponding winding number is
\be
Q = l^2.
\ee

It is worth mentioning that, as shown in \cite{decarli,bonfim}, the solutions \rf{solution1} and \rf{solution2} also solve the model \rf{ua845} in the case when the quadratic term is absent, {\it i.e.} $m=0$,  and in consequence,  the $\vp$ field is also absent.


\section{Conclusions and Final Remarks}

In this paper we have defined a new Skyrme-like model which has an exact BPS bound. The basic formulation of the model does 
not depend on the properties of the space-time manifold over which the model was defined. 

When we have tried to find finite energy solutions of this model we have discovered that its BPS equation, known as force-free equation, 
  has already been used to  describe some phenomena in plasma and solar physics applications. Most of these phenomena
involved fields defined over finite or semi-finite domains and we have found an old observation of Chandrasekhar that the model
does not possess finite energy solutions when considered as a model for smooth fields over the entire $\IR^3$.
This, of course, does not prevent the model from having any compacton-like solutions but, so far, we have not succeeded 
in finding any. 

However, the model does possess finite energy solutions when we consider it over other space-times. So we looked at its
solutions over  $S^3$ and in the last section of the paper we have presented two classes of such solutions. The solutions depend
on one integer parameter $l$ and for any value of this parameter 
 both have the same energy and the same topological charge (proportional to $l^2$).
We are now studying other properties of the model and checking whether the model possesses further solutions.

\vspace{4cm}

\noindent {\bf Acknowledgemnts} LAF is very grateful to Profs. Jean-Jacques Aly and Norbert Seehafer for many helpful e-mail messages explaining the solutions of the force-free equations as well as for providing many references on that subject. LAF and WJZ are very grateful to Dr. Andrzej Wereszczynski for many helpful discussions in the early stages of this work. LAF is partially supported by CNPq-Brazil.
The work reported here was completed when the authors visited the Mathematisches Forschungsintitut in Oberwolfach (MFO). The authors would like to thank the MFO for its hospitality.

\end{document}